\documentclass[aps,showpacs,eqsecnum,twocolumn,superscriptaddress,preprintnumbers]{revtex4}

\usepackage{amsmath}
\usepackage{latexsym}
\usepackage{graphicx}
\graphicspath{{figures/}}

\begin{document}
\preprint{AEI-2005-002}


\title{Excision methods for high resolution shock capturing schemes
applied to general relativistic hydrodynamics}


\author{Ian~Hawke}
\affiliation{Max-Planck-Institut f\"ur Gravitationsphysik,
Albert-Einstein-Institut, 14476 Golm, Germany}
\affiliation{School of Mathematics, University of Southampton,
  Southampton SO17 1BJ, UK}
\author{Frank~L\"offler}
\affiliation{Max-Planck-Institut f\"ur Gravitationsphysik,
Albert-Einstein-Institut, 14476 Golm, Germany}

\author{Andrea~Nerozzi}
\affiliation{Institute of Cosmology and Gravitation,
Mercantile House, Hampshire Terrace, PO1 2EG, Portsmouth UK}
\affiliation{Center for Relativity, University of Texas at
Austin, Austin TX 78712-1081, USA}


\date{\today}


\begin{abstract}
  We present a simple method for applying excision boundary
  conditions for the relativistic Euler equations. This method depends
  on the use of Reconstruction-Evolution methods, a standard class of
  HRSC methods.  We test three different reconstruction schemes, namely
  TVD, PPM and ENO. The method does not require that the coordinate system
  is adapted to the excision boundary.
  We demonstrate the effectiveness of our method using tests containing 
  discontinuites, static test-fluid solutions with black holes, and full 
  dynamical collapse of a neutron star to a black hole.
  A modified PPM scheme is introduced because of
  problems arisen when matching excision with the original PPM
  reconstruction scheme. 
\end{abstract}


\pacs{
02.70.Bf, 
04.25.Dm, 
04.30.Db, 
95.30.Lz, 
97.60.Lf  
}


\maketitle


\section{Introduction}
\label{sec:introduction}

With gravitational wave detectors such as LIGO, VIRGO and GEO
operational the problem of calculating gravitational wave templates
has become even more urgent. Amongst the physical models that are the
best candidates for producing detectable wave signals those including
highly relativistic matter near black holes stand out. In cases such
as black hole / neutron star binaries, binary neutron star systems
that collapse promptly to a black hole, accretion flows onto black
holes and many models for gamma-ray bursts, detailed numerical
simulations will be required to find the impact of varying physical
parameters on the gravitational waves produced.

Black hole / neutron star binaries are on astrophysical grounds believed to be
as likely as binary neutron star mergers, with expected event
rates of one per year in a sphere of about $70$ Mpc
radius~\cite{Bethe1998}.  While signals from binary neutron stars are
expected to give us information about the masses, spins and locations
of the objects, they are not expected to give information about the
internal structure of the stars. Signals from mixed binary systems, on
the other hand, will provide information about the neutron star
structure and equation of state (EOS)~\cite{Vallisneri2002}.

The crucial problem for numerical simulations involving 3D general
relativity which must be overcome to simulate such physical systems is
stability. With current formulations of the vacuum Einstein equations
it is possible to produce long term simulations of black holes in
certain situations~\cite{Alcubierre00a,Alcubierre01a,Yo02a}. These
simulations typically require some part of the computational domain
inside the black hole to be {\em excised}, with an inner boundary
condition placed on a surface inside the apparent horizon. This
apparent horizon is never outside the event horizon. Because no
physical signal can travel outwards from such an horizon, excising the
interior (or parts of it) should not affect the exterior spacetime,
which is the only region we can observe.  The main reason to excise
the part of the spacetime containing the singularity is that otherwise
steep gradients near the physical singularity form, which numerical
codes cannot handle.  The only long term stable simulations including
black holes performed without
excision~\cite{Alcubierre02a,Alcubierre2003:BBH0-excision} use
techniques that are only applicable when the black hole is present in
the initial slice.

In contrast most simulations including hydrodynamics have either been
performed on a fixed spacetime background, or have only been run until
a short time after the formation of the black
hole~\cite{Shibata:2003iw}.  There have been fully dynamical
simulations of matter with black holes in axisymmetry such
as~\cite{Brandt98}, but few in 3D~\cite{Baiotti04,Duez04,Duez04hydro}.

In this paper we will present a simple method for excision boundaries
applied to hydrodynamics. The boundary condition is based on
High--Resolution Shock--Capturing (HRSC) methods which may be used in
a hydrodynamics code, and theoretically could be applied to any system
using such HRSC methods. We show how it can be applied to three
different, standard reconstruction schemes: TVD, ENO and PPM. Because
of some problems found using this boundary condition with the PPM
scheme we introduce a modified version of PPM (MPPM), which solves
these problems.  This excision method, combined with a suitable
excision method for the spacetime, allows long term simulations of
matter in black hole spacetimes. This has been shown e.g.\ using our
hydrodynamics code called {\tt Whisky} in~\cite{Baiotti04}.

The outline of this paper is as follows. In section~\ref{sec:model}
and~\ref{sec:method} respectively we outline the equations
and HRSC methods that will be used. The modifications required at
excision boundaries are given in section~\ref{sec:excision}.
Section~\ref{sec:tests} contains the tests used to validate the
boundary conditions.  Throughout this paper we shall use geometric
units where $c = G = M_{\odot} = 1$. Greek indices are taken to run
from 0 to 3, latin indices from 1 to 3. We adopt the standard
convention for the summation over repeated indices. In
section~\ref{sec:method} latin indices denote the cell index.


\section{Model and equations}
\label{sec:model}

We are interested in simulating hydrodynamical flows near black holes,
so we use the equations of general relativistic hydrodynamics coupled to
a dynamical spacetime, described by full general relativity.
We use the flux-conservative Valencia 3+1
formulation of the hydrodynamical equations \cite{Marti91, Banyuls97,
Ibanez01}.
Although the results of Lax and Wendroff~\cite{Lax60} show
that if the scheme converges then it will converge to one of the
(possibly infinitely many) weak solutions of the system of equations,
the use of the correct conservation law form is known to be
crucial as the results of Hou and Lefloch~\cite{Hou94} show that a
non-conservative scheme will \textit{generically} converge to the
wrong weak solution.
The stress energy tensor is written as
\begin{equation}
  \label{eq:Tmunu}
  T_{\mu\nu} = \rho h u_{\mu}u_{\nu} + p g_{\mu\nu},
\end{equation}
where $\rho$ is the mass density of the fluid, $h=1 + \epsilon +
p/\rho$ the specific enthalpy with $\epsilon$ the specific internal
energy, and $p$ the pressure. $u_{\mu}$ is the 4-velocity of the fluid
and $g_{\mu\nu}$ the 4-metric of the spacetime.

The equations of relativistic hydrodynamics can then be written in a
conserved form
\begin{equation}
  \label{eq:conservedeqn}
  \partial_t {\bf q} + \partial_{j} {\bf f}^{(j)}({\bf q}) = {\bf
  s}({\bf q}).
\end{equation}
Here we use the Valencia form \cite{Marti91,Banyuls97} as in
\cite{Font00b,Font02c,Baiotti04}. The ${\bf q}$ is a vector of
conserved variables. They are not strictly conserved because of the
source terms ${\bf s}$, but in flat space the sources ${\bf s}$
vanish.

The system requires an EOS to close it.
This is normally given by specifying the pressure $p$ in the form
$p=p(\rho, \epsilon)$. In this paper we will consider the standard
perfect fluid gamma-law EOS
\begin{equation}
  \label{eq:ifeos}
  p = (\Gamma - 1)\rho\epsilon,
\end{equation}
and often we will restrict to the polytropic EOS
\begin{equation}
  \label{eq:polyeos}
  p = K \rho^{\Gamma},
\end{equation}
which is a good approximation to neutron star matter in a cold neutron
star and in the absence of shocks.


\section{Numerical methods}
\label{sec:method}

The numerical methods that we use for the evolutions of the hydrodynamic
variables are all High--Resolution
Shock--Capturing (HRSC) methods. We take the semidiscrete or method of
lines reconstruction-evolution viewpoint: a piecewise continuous
interpolant (the \textit{reconstruction}) of each variable is found,
at each cell boundary a Riemann problem is solved (usually
approximately), the time derivative for each variable is constructed
from the flux differences through the boundaries of each cell and the
source terms are calculated, and the solution at the new timelevel is
found using some suitable time integrator.

In what follows we shall specialise to the case of a uniform Cartesian
grid.

\subsection{Riemann solvers}
\label{sec:riemann-solv}

Our code implements three independent approximate Riemann solvers
(HLLE, Roe, modified Marquina). Although an integral part of the full
HRSC method they are irrelevant for the problem of excision, as
explained below.  The Riemann solver employed in all tests shown below
is the modified Marquina solver described in \cite{Aloy99a,Aloy99b}.

\subsection{Reconstruction methods}
\label{sec:reconstr-meth}

Four separate reconstruction methods are considered here. Each are
applied in a dimensionally split fashion; the reconstructions are
performed along each coordinate axis in turn. Although some of the
reconstruction methods considered here have formal orders of accuracy
that are better than second order, the formal global order of accuracy
of the code is at best second order. This is due to the extension to
multiple dimensions and the coupling to the spacetime. The flux
through the cell boundary is approximated by the flux through the
point in the middle of the cell boundary, using the reconstructed
fluid variables and a second order approximation of the metric terms.
The use of reconstruction methods that have one dimensional formal
orders of accuracy better than second order leads to improvements in
actual accuracy but not in the formal convergence order.

The method with the lowest formal order of accuracy is a slope limited
total variation diminishing (TVD) method. The essentially
non-oscillatory (ENO) methods may have extremely high orders of
accuracy and are more accurate in absolute terms than TVD. However, it
is often more efficient to use the piecewise-parabolic method (PPM)
even though it only has at most third--order accuracy. The fourth
method is a variant of PPM, and it is introduced for the first time
here. It solves some problems found using PPM
near the excision region.  Therefore we call it MPPM for `modified
PPM'.

For conventions about subscripts and superscripts to denote cells,
cell boundaries and left or right sided cell boundary values also see
Fig.~\ref{fig:hydroexcise}.

\subsubsection{Slope--limited TVD}
\label{sec:tvd}

Total variation diminishing (TVD) methods as given by,
e.g.,~\cite{vanLeer79}, ensure that the solution remains monotonic to
avoid spurious oscillations near discontinuities. In the case of
slope-limited TVD the reconstruction is given by an average of a
first--order and a second--order reconstruction. To reconstruct the
variable $q$ in the cell centred at $x_i$ two local ``slopes'' are
defined,
\begin{equation}
  \label{eq:TVD1}
  \begin{array}[c]{r c l}
    \Delta^-_i & \equiv & q_{i} - q_{i-1}, \\
    \Delta^+_i & \equiv & q_{i+1} - q_{i}, 
  \end{array}
\end{equation}
which are then averaged to get the slope in the cell $\bar{\Delta}_i =
(\Delta^-_i+\Delta^+_i)/2$. This slope is then multiplied by a {\em
  limiter} which is a function $\phi$ of the local slopes $\phi =
\phi(\Delta^-_i,\Delta^+_i)$. This {\em limited slope}
$\bar{\Delta}_i$ then gives the cell boundary data by
\begin{equation}
  \label{eq:TVD2}
  q_{i \pm 1/2} = q_i \pm \frac{1}{2} \bar{\Delta}_i.
\end{equation}

If the slopes are not limited (i.e., the limiter $\phi = 1$) then the
method is second order accurate. However, in the presence of steep
gradients or discontinuities such a reconstruction will be oscillatory
due to Gibb's effects. In these regions the limiter reduces the slopes
to avoid overshoots, retaining monotonicity of the reconstruction.

\subsubsection{ENO}
\label{sec:eno}

The ENO methods of Harten et
al.~\cite{Harten87} are a very general class of methods. Here we only
consider the simple ENO reconstruction of the variables as given by
Shu in~\cite{Shu99}. This provides arbitrary order of accuracy in
space.

As in the TVD case, ENO reconstructions are cell based. The
$p^{\text{th}}$ order reconstruction compares all possible polynomial
stencils of size $p$ containing the cell to be reconstructed. The
``least oscillatory'' stencil is chosen by minimizing the absolute
values of the divided differences that make up the stencil.

Let $p$ be the order of the reconstruction. Suppose we are
reconstructing the scalar function $q$ in cell $i$. We start with the
cell $i$. We then add to the stencil cell $j$, where $j = i \pm 1$,
where we choose $j$ to minimize the Newton undivided differences
\begin{subequations}
  \label{eq:enodd}
  \begin{eqnarray}
    q \left[ i-1, i \right] & \equiv & q_i - q_{i-1}, \\
    q \left[ i, i+1 \right] & \equiv & q_{i+1} - q_i.
  \end{eqnarray}  
\end{subequations}
We then recursively add more cells, minimizing the higher order Newton
divided differences $q \left[ i-1, \dots, i+j \right]$ defined by
\begin{equation}
  \label{eq:enodd2}
  \begin{array}[c]{r c l}
  q \left[ i-1, \dots, i+j \right] &= &q \left[ i, \dots, i+j \right] -\\
                                     &&q \left[ i-1, \dots, i+j-1 \right].
  \end{array}
\end{equation}
The reconstruction at the cell boundary is given by a standard
$p^{\textrm{th}}$ order polynomial interpolation on the chosen
stencil.

Shu~\cite{Shu99} has outlined an elegant way of calculating the cell
boundary values solely in terms of the stencil and the known data. If
the stencil is given by
\begin{equation}
  \label{eq:enostencil1}
  S(i) = \left\{ i-r, \dots, i+p-r-1 \right\},
\end{equation}
for some integer $r$, then there exist constants $c_{rj}$ depending
only on the grid $x_i$ such that the boundary values for cell $I_i$
are given by
\begin{equation}
  \label{eq:enoc1}
  \begin{array}[c]{r c l}
  q_{i+1/2} &= &\sum_{j=0}^{p-1} c_{r,j} q_{i-r+j},\\
  q_{i-1/2} &= &\sum_{j=0}^{p-1} c_{r-1,j} q_{i-r+j}. 
  \end{array}
\end{equation}
The constants $c_{r,j}$ are given by the rather complicated formula
\begin{widetext}
\begin{equation}
  \label{eq:enoc2}
  c_{r,j} = \left\{ \sum_{m=j+1}^p \frac{ \sum_{l=0, l \neq m}^p
  \prod_{q=0, q \neq m, l}^p \left(  x_{i+1/2} - x_{i-r+q-1/2} \right)
  }{ \prod_{l=0, l \neq m}^p \left(  x_{i-r+m-1/2} - x_{i-r+L-1/2}
  \right) }  \right\} \Delta x_{i-r+j}. 
\end{equation}
\end{widetext}
This calculation simplifies considerably if the grid is evenly spaced. For
this case the coefficients up to seventh--order are given by
Shu~\cite{Shu99}.

\subsubsection{PPM}
\label{sec:ppm}

The Piecewise Parabolic Method (PPM) of Colella and
Woodward~\cite{Colella84}, generalized to relativistic flows by
Mart\'\i~and M\"uller~\cite{Marti96}, is third--order accurate in space
with particular special cases to ensure monotonicity at shocks,
sharpening of contact discontinuities, and shock detection.

The outline of the general PPM method is as follows. The first step is
to interpolate a quartic polynomial to the cell boundary,
\begin{equation}
  \label{eq:ppm1}
  q_{i+1/2} = \frac{1}{2} \left( q_{i+1} + q_i \right) + \frac{1}{6}
  \left( \delta_m q_i - \delta_m q_{i+1} \right), 
\end{equation}
where
\begin{widetext}
\begin{equation}
  \label{eq:ppmdm1}
  \delta_m q_i = \left\{ \begin{array}{c l} \textrm{min}(|q_{i+1} - q_{i-1}|,
      2|q_{i+1} - q_i|, 2|q_i - q_{i-1}|) \textrm{ sign}(q_{i+1} - q_{i-1}) &
      \textrm{   if } (q_{i+1} - q_i)(q_i - q_{i-1}) > 0 \\*[1em]
      0 & \textrm{   otherwise} \end{array} \right..
\end{equation}
\end{widetext}
At this point we set both left and right states at the interface to be
equal to the interpolated value,
\begin{equation}
  \label{eq:ppmset1}
  q_{i+1/2}^+ = q_{i+1/2}^- = q_{i+1/2}.
\end{equation}

This reconstruction will be oscillatory near shocks. Before a step
which will preserve the monotonicity there are two other steps, which
may be applied.

Firstly we may ``steepen'' discontinuities. This is to produce sharper
profiles and is only applied to discontinuities that are mostly a contact
(see~\cite{Colella84} for the details).  This procedure
replaces the cell boundary reconstructions of the density with
\begin{widetext}
\begin{subequations}
  \label{eq:ppmdetect}
  \begin{eqnarray}
    \rho_{i-1/2}^+ & \equiv & \rho_{i-1/2}^+ (1-\eta) + \left(\rho_{i-1} +
      \frac{1}{2} \delta_m \rho_{i-1} \right) \eta, \\
    \rho_{i+1/2}^- & \equiv & \rho_{i+1/2}^- (1-\eta) + \left(\rho_{i+1} -
      \frac{1}{2} \delta_m \rho_{i+1} \right) \eta, 
  \end{eqnarray}
\end{subequations}
where $\eta$ is defined as
\begin{equation}
  \label{eq:ppmeta}
  \eta \equiv \textrm{max}\left[0, \textrm{min}
              \left(1, \eta_1 (\tilde{\eta} - \eta_2)\right)\right],
\end{equation}
where $\eta_1,\eta_2$ are constants and
\begin{equation}
  \label{eq:ppmetatilde}
  \tilde{\eta} \equiv \left\{ \begin{array}{c l} -\dfrac{1}{6}
  \dfrac{\rho_{i+2} - 2 \rho_{i+1} + 2 \rho_{i-1} - \rho_{i-2}}
       {\rho_{i+1} - \rho_{i-1}} &
  \textrm{ if } -\delta^2\rho_{i+1}\delta^2\rho_{i-1} > 0\mbox{ and }
                \left|\rho_{i+1} - \rho_{i-1}\right| -
                  \epsilon_S \textrm{ min}(|\rho_{i+1}|,|\rho_{i-1}|) > 0
  \\*[1em]
  0 & \textrm{otherwise} \end{array} \right.
\end{equation}
with $\epsilon_S$ another constant and
\begin{equation}
  \label{eq:ppmd2rho}
  \delta^2\rho_i \equiv \rho_{i+1} - 2\rho_i + \rho_{i-1}.
\end{equation}
\end{widetext}
Suggested values for the constants $\eta_1,\eta_2$ and $\epsilon_S$ can be found
in~\cite{Colella84}.

Another step that may be performed before monotonicity enforcement is the
``flattening'' of the zone structure near shocks. This adds simple
dissipation, altering the reconstructions to
\begin{widetext}
\begin{eqnarray}
  \label{eq:ppmflatten}
  q_{i-1/2}^+ & \equiv & \nu_i q_{i-1/2}^+ + (1 - \nu_i) q_i, \\
  q_{i+1/2}^- & \equiv & \nu_i q_{i+1/2}^- + (1 - \nu_i) q_i,
\end{eqnarray}
where
\begin{equation}
  \label{eq:ppmflatten2}
  \nu_i \equiv \left\{ \begin{array}{c l} \textrm{max}\left[0, 
    1 - \textrm{max}\left(0,
      \omega_2 \left(
         \dfrac{p_{i+1} - p_{i-1}}{p_{i+2} - p_{i-2}} - \omega_1\right)\right)
    \right] & \textrm{
        if  } \epsilon \textrm{ min}(p_{i-1}, p_{i+1}) < p_{i+1} - p_{i
        - 1} \textrm{ and } v^x_{i-1} - v^x_{i+1} > 0 \\*[1em]
      1 & \textrm{otherwise} \end{array} \right.,
\end{equation}
\end{widetext}
and $\omega_1,\omega_2$ and $\epsilon$ are again constants with suggested values
given in~\cite{Colella84}. Note that this step is a
simplification of the one in~\cite{Colella84,Marti96}, which reduces
the communication overhead in parallel runs. No problems were
encountered using this modification.

The final step is applied to preserve monotonicity, which prevents
oscillatory reconstruction near shocks. The following replacements are
made:
\begin{widetext}
\begin{subequations}
  \label{eq:ppmmonot}
  \begin{eqnarray}
    q_{i-1/2}^+ \equiv q_{i+1/2}^- = q_i & \textrm{   if} & (q_{i+1/2}^- -
    q_i)(q_i - q_{i+1/2}^-) \leq 
    0, \\
    q_{i-1/2}^+ \equiv 3 q_i - 2q_{i+1/2}^- & \textrm{   if} & (q_{i+1/2}^- -
    q_{i-1/2}^+)\left( q_i - \frac{1}{2} (q_{i-1/2}^+ + q_{i+1/2}^-)
    \right) > \frac{1}{6}(q_{i+1/2}^- - q_{i-1/2}^+)^2,
    \\ 
    q_{i+1/2}^- \equiv 3 q_i - 2q_{i-1/2}^+ & \textrm{   if} & (q_{i+1/2}^- -
    q_{i-1/2}^+)\left( q_i - \frac{1}{2} (q_{i-1/2}^+ + q_{i+1/2}^-)
    \right) < -\frac{1}{6}(q_{i+1/2}^- - q_{i-1/2}^+)^2. 
  \end{eqnarray}  
\end{subequations}
\end{widetext}

\subsubsection{MPPM}
\label{sec:mppm}

As we shall see below there are situations where the PPM scheme does
not give good results. This is due to the use of a fixed stencil in
the first part of the PPM algorithm where the interpolated value is
computed. This can cause problems with supersonic flows and when, as
in our excision method, we try to match to a non-PPM method. To avoid
this problem the modified PPM scheme changes the interpolated value in
\eqref{eq:ppm1}.  The modified method uses the same steps as the PPM
scheme after the interface values are set in Eq.~\eqref{eq:ppmset1}
(discontinuity steepening, zone flattening and monotonicity preservation).

PPM uses the points at $i-1$, $i$, $i+1$, $i+2$ for the fourth order
interpolation function to obtain a value at $i+\frac{1}{2}$. In the
case of supersonic flow this centred stencil takes too much
information from one side and too little from the other. Because of
this, the method may produce small amplitude oscillations that are
stable (due to the other steps in the PPM method) but which do not
converge away. To cure this we need to use a stencil that depends on
the data in some way.

In particular, we calculate the minimum and maximum characteristic
velocities from the minimum and maximum eigenvalues $\lambda_-$ and
$\lambda_+$ of the Jacobian matrix $\partial {\bf f} / \partial {\bf
  q}$.  Depending on these values, we define
\begin{equation}
 \label{eq:mppmalpha}
 \alpha=\frac{\lambda_-+\lambda_+}
             {\left|\lambda_-\right|+\left|\lambda_+\right|},
\end{equation}
which is a continuous function of the characteristic speeds that has
value $\pm 1$ whenever the flow is supersonic. Because the eigenvalues
can only be in the range $[-1,1]$ $\alpha$ is also in that range.

Using $\alpha$ we calculate the initial boundary states
\begin{equation}
 \label{eq:mppminital}
 q_{i+1/2}^+ = \left\{ \begin{array}{c c l}
                \left|\alpha\right| q_L +
                \left(1-\left|\alpha\right|\right) q_{i+1/2}, & \textrm{   if}
                & \alpha < 0, \\
               \left|\alpha\right| q_R +
                \left(1-\left|\alpha\right|\right) q_{i+1/2}, & \textrm{   if}
                & \alpha > 0 \end{array} \right.
\end{equation}
with
\begin{eqnarray}
 \label{eq:mppmdefs}
 q_L &=& \frac{1}{12}\left(13q_{i+1}-5q_{i+2}+q_{i+3}+3q_i    \right),\\
 q_R &=& \frac{1}{12}\left(13q_i    -5q_{i-1}+q_{i-2}+3q_{i+1}\right).
\end{eqnarray}
In this way we shift the stencil at most one cell to the side
depending smoothly on the data. Note that this choice is arbitrary and
could be chosen in a different way, but was sufficient for the tests
used here.  Because the value of the interpolating function at the
boundary can be outside the range given by the values at the
neighbouring cells $q_i$ and $q_{i+1}$, we set the boundary values
$q_{i+1/2}^+$ and $q_{i+1/2}^-$ in that case to be equal to the
closest nearby cell value.  Once the interpolated value is set we
proceed as in the PPM case with the steps required to, e.g., preserve
monotonicity, as given in
Eqs.~(\ref{eq:ppmdetect}--\ref{eq:ppmmonot}).

This procedure takes the characteristic speeds of the data into
account and should therefore be more suitable to extreme cases where
both minimal and maximal speeds have the same sign and are not close
to zero. However, in cases with minimal and maximal characteristic
speeds of the same size, but different sign, it should be as good as
PPM.

\section{Excision boundaries}
\label{sec:excision}

\setcounter{subsubsection}{0}

Excision boundaries are based on the principle that a closed region of
spacetime that is causally disconnected from the rest of the simulated
spacetime can be ignored without affecting the results in the exterior
spacetime. A black hole event horizon is given indeed by the
boundary of such a causally disconnected region of spacetime. As
generically a curvature singularity will form within the horizon it
will be necessary to remove it, and the associated steep gradients,
from the numerical domain. 

\begin{figure*}[htbp]
  \begin{center}
    \includegraphics*[angle=0,scale=0.5]{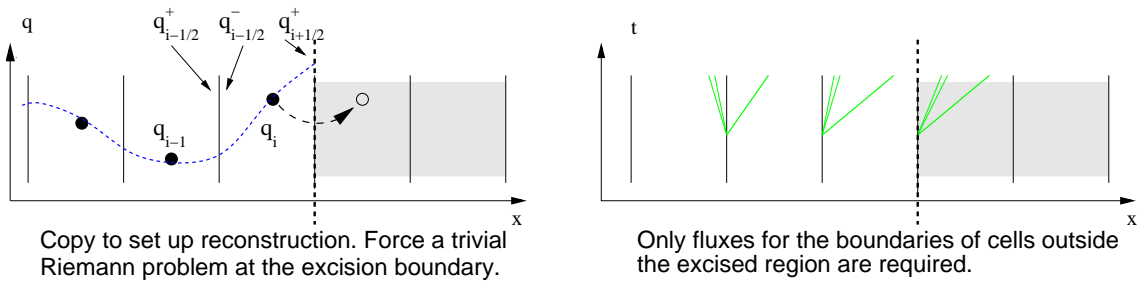}
    \caption{A schematic view of the excision algorithm.  The excised
      region is indicated by the shaded grey cells and the excision
      boundary by the vertical dotted line. In the left panel we
      indicate how the reconstruction method is modified near the
      excision boundary. The values required to compute the fluxes for
      the cell next to the boundary are indicated, along with the
      method used (simple copying) to ``extend'' the data across the
      excision boundary. Using this we can compute the fluxes through
      the cell boundaries for all cells outside the excised region. As
      shown in the right panel, these are the only fluxes required.}
    \label{fig:hydroexcise}
  \end{center}
\end{figure*}

Excision boundaries are usually placed within a trapped surface such
as an apparent horizon, as it is not possible to find the event
horizon locally in time. We note that for a cubical region that is
excised on a Cartesian grid to be a true trapped surface it may have
to be placed well within the horizon, as pointed out
by~\cite{Calabrese:2003a}. However, inflow boundary conditions applied
to surfaces within the apparent horizon have been shown to work
well~\cite{Alcubierre2003:BBH0-excision}.

The same considerations apply to the hydrodynamical variables as to
the spacetime field variables. Again there will be a region of
spacetime in which the fluid is causally disconnected from the
exterior. In practise due to the coupling with the spacetime field
this will coincide with the event horizon as for the spacetime
field. However, if the spacetime field variables have been excised
then in principle there is no need to excise the hydrodynamical field
variables at all, as these are not expected to blow up and form
singularities independently of the spacetime field variables.

In practise unacceptable numerical errors are found if the
hydrodynamical variables are not excised. In particular at typical
resolutions it is possible for numerical effects in the interior of
the black hole horizon to propagate out of the horizon through the
hydrodynamical variables. Although we will exploit the independence
{\em in principle} of the use of excision boundaries for the different
field variables later, in practise both should be used simultaneously.

Excision boundary methods can be viewed in a number of different
ways. Viewed at the level of the differential equations, all
characteristics of physical quantities
are going into the excision region. At the level of
the Riemann problem, all possible waves from the solution of the
Riemann problem must be contained within the excision region.

This immediately suggests a method that could be used at the excision boundary.
As a HRSC method naturally changes the stencil locally depending on the
data, we just need to guarantee that the data used to construct the
flux at the excision boundary is the physically relevant data outside
the excision region. This also suggests that the natural choice to put
the excision boundary is not a cell centre, but a cell boundary.

While very simple, this method is not stable in general. It is
equivalent to filling the first cell inside the excision region by
linear extrapolation from the exterior (for a second--order HRSC
method such as slope limited TVD reconstruction). This is not
guaranteed to reduce the total variation of the solution, and so even
simple examples do fail with this boundary condition.

On the other hand the simplest outflow boundary condition at the
excision boundary does not have this problem. In particular, we may
apply zero--order extrapolation (a simple copy) to all variables at
the boundary to create data at the first cell inside the excision
region. This is done for each of the three reconstruction methods.
If one method requires more cells for the
stencil in the interior of the excision region, we can then force the
stencil to only consider the data in the first cell and the exterior
as above.

A summary of our method is as follows:
\begin{enumerate}
  \item Any point for which all possible reconstruction stencils are
  contained within the exterior spacetime (i.e., do not contain an
  excised point) is evolved in the normal fashion.
  \item Any point which is contained within the excised region is not
  evolved at all.
  \item All other points have at least one stencil that contains an
  excision point. To update these points we must reconstruct the data
  to both sides of each cell boundary. Once this is done the Riemann
  problem can be solved, the boundary flux found and the update
  computed. In order to reconstruct the data we do the following:
  \begin{enumerate}
    \item The values for all variables in the cell next to the
      excision boundary are copied to the first cell within the excision
      boundary. 
    \item A reconstruction of all cells is produced using only the
      data outside the excision region and the first cell within. The
      precise method depends on the global reconstruction method used
      and is explained below. In outline, the idea is to retain the
      order of accuracy by retaining the same size of the stencil if
      possible.
    \item This method produces reconstructed data on both sides of
      every cell boundary {\em except} for the excision boundary,
      where only the reconstruction from the exterior can be
      produced. At this boundary we assume on physical grounds that
      the solution of the Riemann problem must be given by the
      exterior state (as this is assumed to be an outflow
      boundary). Therefore at these points we do not solve the Riemann
      problem, but calculate the flux from the exterior reconstruction
      (see Fig.~\ref{fig:hydroexcise}).
  \end{enumerate}
\end{enumerate}



As an aside, we note that this means the method is independent of the
choice of Riemann solver. However, the choice of a {\em trivial}
Riemann problem at points that are not excised but where the
reconstructed states are identical (e.g., in the atmosphere) can lead
to gains in computational efficiency.

To completely describe our method in practice we just need to describe
how the stencil is altered with the specific reconstruction methods we
use. In what follows we consider a set of cells in one dimension. The
coordinate is labelled $x$, the cell centres $x_i$, and the cell
boundaries $x_{i \pm 1/2}$ in an obvious notation. The excision
boundary will be denoted $x_{B + 1/2}$, with the cell $x_{B+1}$ being
excised. We want to calculate the update term for the cell centred at
$x_B$ by calculating the fluxes at $x_{B \pm 1/2}$.

Excision of the spacetime variables is applied using the
{\it simple lego} excision method described in,
e.g.,~\cite{Alcubierre2003:BBH0-excision,Alcubierre00a}.

\subsubsection{Slope-limited TVD}
\label{sec:excision-tvd}

In this case it is clear that only the reconstructions at $x_{B \pm
  1/2}$ are affected by the excision region. It is also clear that
setting
\begin{equation}
  \label{eq:TVD3}
  \bar{\Delta}_B = \frac{1}{2}\left(\Delta^-_B+\Delta^+_B\right)=0,
\end{equation}
ensures that only data outside the excision region is used and is
consistent with the TVD reconstruction.

\subsubsection{ENO}
\label{sec:excision-eno}

As described in section \ref{sec:eno} the ENO method uses a stencil
width depending on the desired order of accuracy $p$. Hence the
reconstruction in the cells centred between $x_B$ and $x_{B - p + 2}$
are affected by the excision region.

It is clear how to ensure that the stencil chosen by the ENO
reconstruction does not include any points from inside the excision
region. The first divided differences that include points within the
excision region are set to extremely large values with oscillating and
growing amplitude such as $(-i)^i \times 10^{10}$ where $i$ is the
index of the cell ($i>0$).  The higher order divided differences are
calculated from these, ensuring that the least oscillatory stencil
does not include points from the excision region, but more points
away from it. This is a simple
application of the principle outlined by Shu~\cite{Shu99} for dealing
with outflow boundaries with ENO methods.

\subsubsection{PPM}
\label{sec:excision-ppm}

PPM uses a point stencil of five points, but reconstructs around a cell boundary
instead of in a cell. Hence the reconstructions affected by the
excision region are the reconstruction at left edge of $x_{B + 1/2}$
and both reconstructions at $x_{B - 1/2}$.

We have not attempted to find a consistent third--order reconstruction
for these points. Instead we use the identical reconstruction as in
the TVD case of section~\ref{sec:tvd}. As this provides a
second--order stable reconstruction at an outflow boundary it does not
affect the order of accuracy in the rest of the domain.

\subsubsection{MPPM}
\label{sec:excision-mppm}

Since excision should only be applied on a boundary with only outgoing
characteristics, $\alpha$ should only be $-1$ or $1$ depending on the
direction of the boundary. Because of the symmetry of the cases, without
loss of generality we
assume $\alpha=1$ from here on, that is assuming the excision boundary
to be to the right of the computational domain.

In the case $\alpha=1$ we effectively use a stencil with one point to
the right and three points to the left instead of the usual symmetric
four point stencil of the interpolation step of PPM.  Because of this,
we can use the usual MPPM
procedure in the whole domain except for only one point next to the
excision boundary.  There we set the left and right boundary values to
be the cell value as in the TVD case.

\section{Numerical tests}
\label{sec:tests}

In this Section we present some simple numerical tests to validate our
approach.
The results were calculated using the Whisky code~\cite{Baiotti04}. Where the 
spacetime was evolved an implementation~\cite{Alcubierre99d,Alcubierre99e}
of the NOK (BSSN)
formulation of~\cite{Shibata95,Baumgarte99} was used, with the excision
method given in~\cite{Alcubierre00a}.
Apparent horizons were found using the
code described in~\cite{Thornburg2003:AH-finding} and event horizons
using the code described in~\cite{Diener03a}.
All codes use the Cactus framework~\cite{Cactusweb}.

\subsection{Shock wave tests}
\label{sec:shocks}

Special relativistic shock tubes are the simplest possible test. For
this test the spacetime background is fixed to be Minkowski spacetime
in standard coordinates. It is simple to choose initial data such that
the chosen excision boundary is an outflow boundary for all time.  In
particular, we are using initial data in which $\rho_L=10$,
$\rho_R=1$, $v_L=v_R=0$, $p_L=40/3$ and $p_R=2/3 \times 10^{-6}$ and
the velocity is normal to the interface.  The ideal fluid equation of
state, Eq.~\eqref{eq:ifeos}, is used, with $\Gamma=5/3$.

We have considered three different cases. The first and simplest was a
shock aligned with a coordinate axis with the excision boundary normal
to the shock. The second and numerically more complex was a shock
along the diagonal of the 3D box with the excision boundary normal to
the shock. The final test consisted of a shock not aligned with the
grid and with the excision boundary given by a sphere.  The test
requires more care in setting the domain evolved as the excision
boundary is no longer an outflow boundary over the entire spacetime,
but only within a localized region. Here we set the domain such that
the intersection of the excision boundary with the domain forms a
hemisphere.
\begin{figure*}[htbp]
  \begin{center}
    \includegraphics*[scale=0.6]{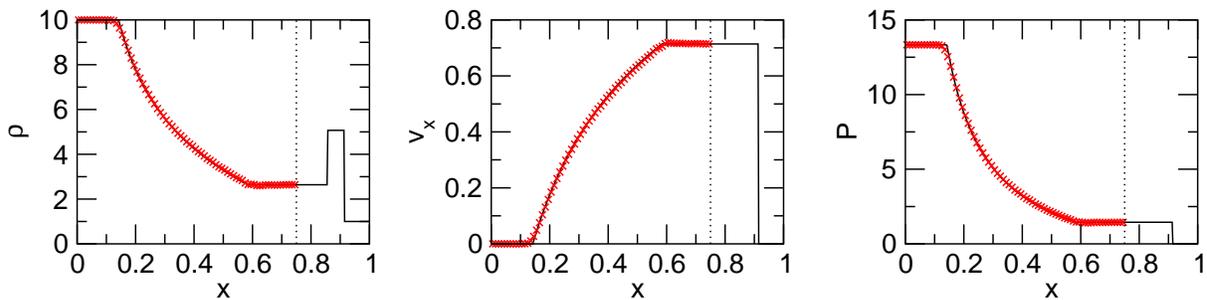}
    \caption{A simple shock propagating parallel to the $x$ axis hits
      an excision boundary normal to the $x$ axis (dotted line).  As
      expected, the shock is completely absorbed. The solid line is
      the exact solution, the crosses are the numerical solution for
      the PPM scheme.  Plots of the other reconstruction methods show
      qualitatively the same behaviour.  The time shown is $t=0.5$.}
    \label{fig:all_mq_shock}
  \end{center}
\end{figure*}

The results of the first test for PPM are shown as an example in
Fig.~\ref{fig:all_mq_shock}. All methods give nearly the same results,
are stable and for all methods the shock passes cleanly through the excision
boundary,
with the results matching the analytic solution well. All other tests
showed similar results, with stable and accurate results independent
of the relative geometry of the shock and excision boundary.

\begin{figure*}[htbp]
  \begin{center}
    \includegraphics*[scale=0.6]{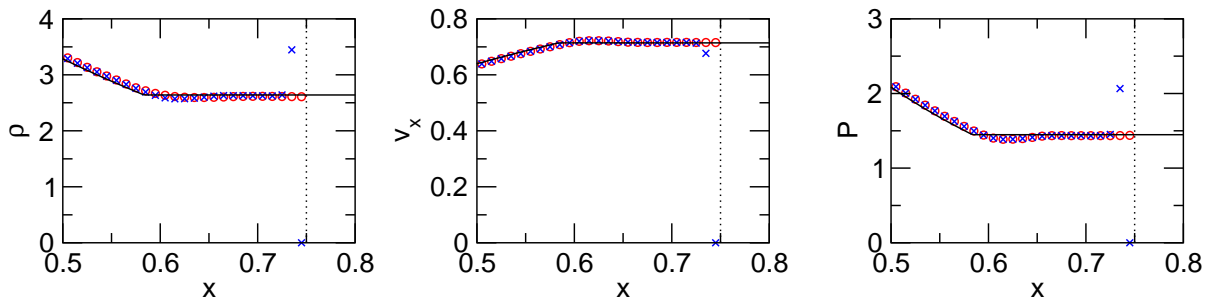}
    \caption{As in Fig.~\ref{fig:all_mq_shock}, a shock is propagated
      parallel to the $x$ axis through an excision boundary. In this
      case a non-conservative excision boundary condition is used.
      This excision boundary condition, which is based on setting all
      variables in the first excised cell to very small
      (``atmosphere'') values gives inaccurate but stable results. The
      circles represent the stable and accurate numerical result found
      using the excision boundary method described in this paper.}
    \label{fig:incorrect_shock}
  \end{center}
\end{figure*}

An example of what may go wrong with an incorrect excision boundary
condition is shown in Fig.~\ref{fig:incorrect_shock}. Here the
standard excision boundary condition presented in
section~\ref{sec:excision} is compared to an alternative method where
the first point inside the excision region is set to very small
values. The result is a stable evolution which is correct except for a
few points near the boundary. Here the density overshoots except for
the point nearest the boundary which itself drops to atmosphere
values. Although in this simple test the boundary condition is
inaccurate, but stable, this seems unlikely to remain the case in more
complex situations. Other boundary conditions, such as first--order
extrapolation to the first point within the excision region, fare even
worse, as they actually fail to produce a stable evolution even on
this simple test.

\subsection{Michel solution}
\label{sec:michel}

The Michel solution \cite{Michel72} is a stationary solution of
spherical accretion onto a Schwarzschild black hole in the test fluid
approximation.  Here we will consider Eddington-Finkelstein
coordinates so that the slice penetrates the horizon. This solution is
often used to validate relativistic hydrodynamic codes
\cite{Papadopoulos98b,Font98a}.

The solution is derived from the basic equations of mass and energy
conservation for the fluid and is obtained numerically using a
Newton--Raphson interation scheme.

\begin{figure*}[htbp]
  \begin{center}
    \includegraphics*[scale=0.6]{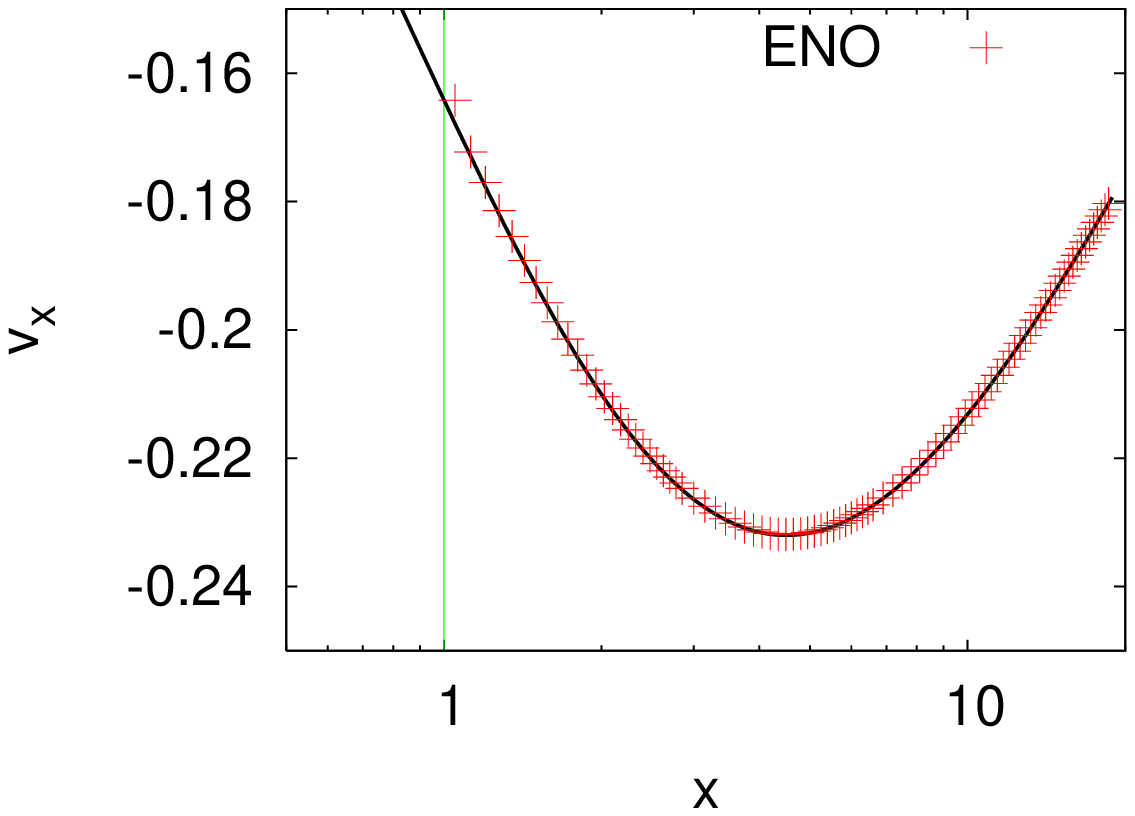}
    \includegraphics*[scale=0.6]{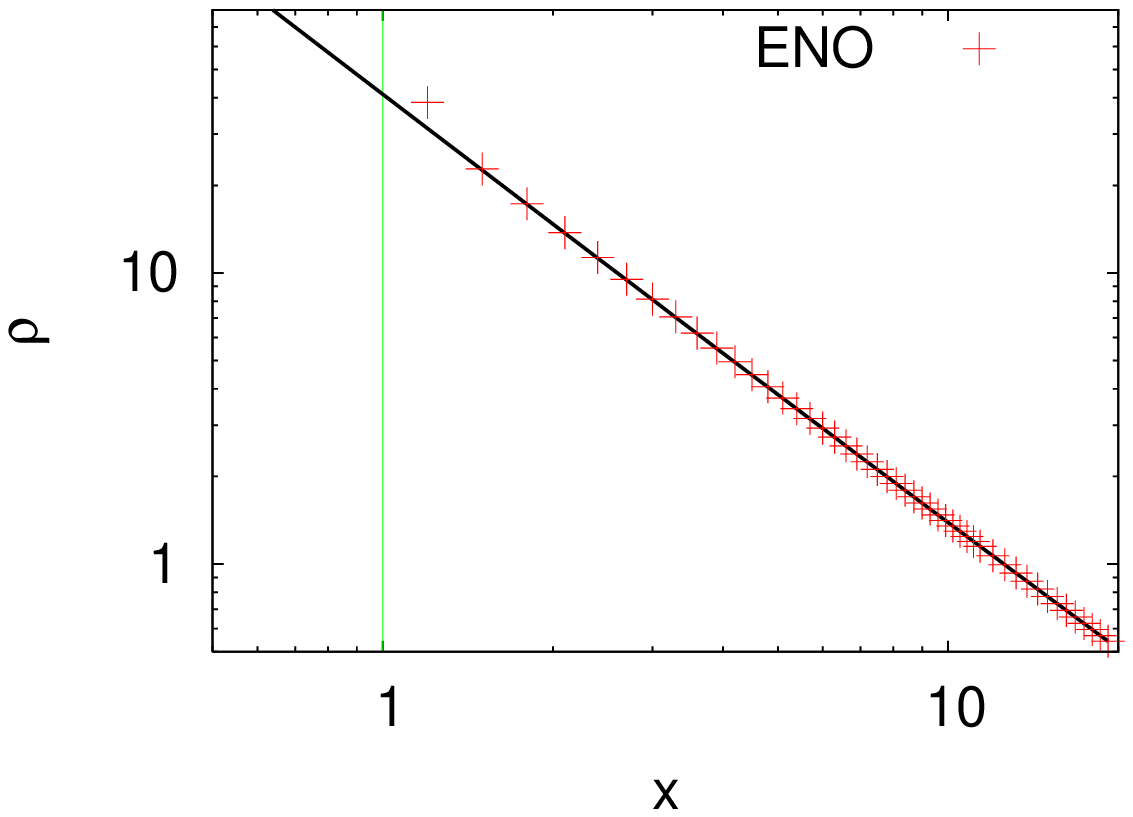}\\
    \includegraphics*[scale=0.6]{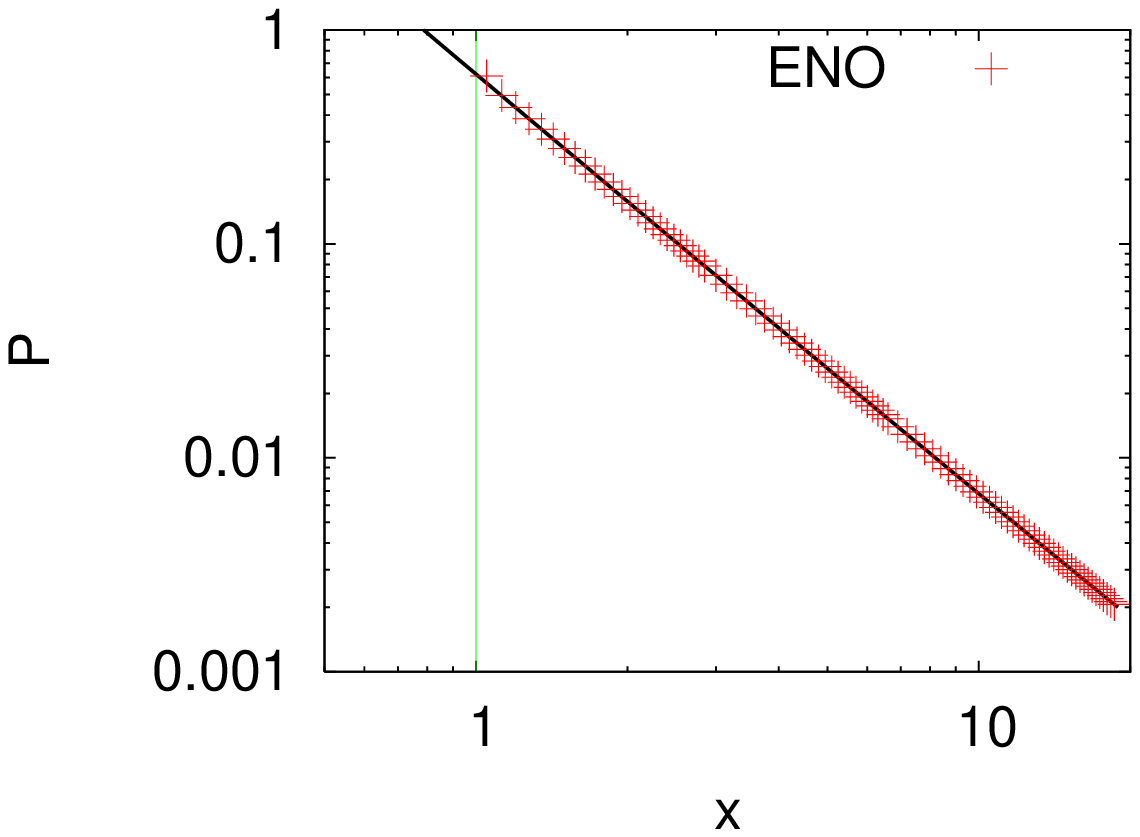}
    \includegraphics*[scale=0.6]{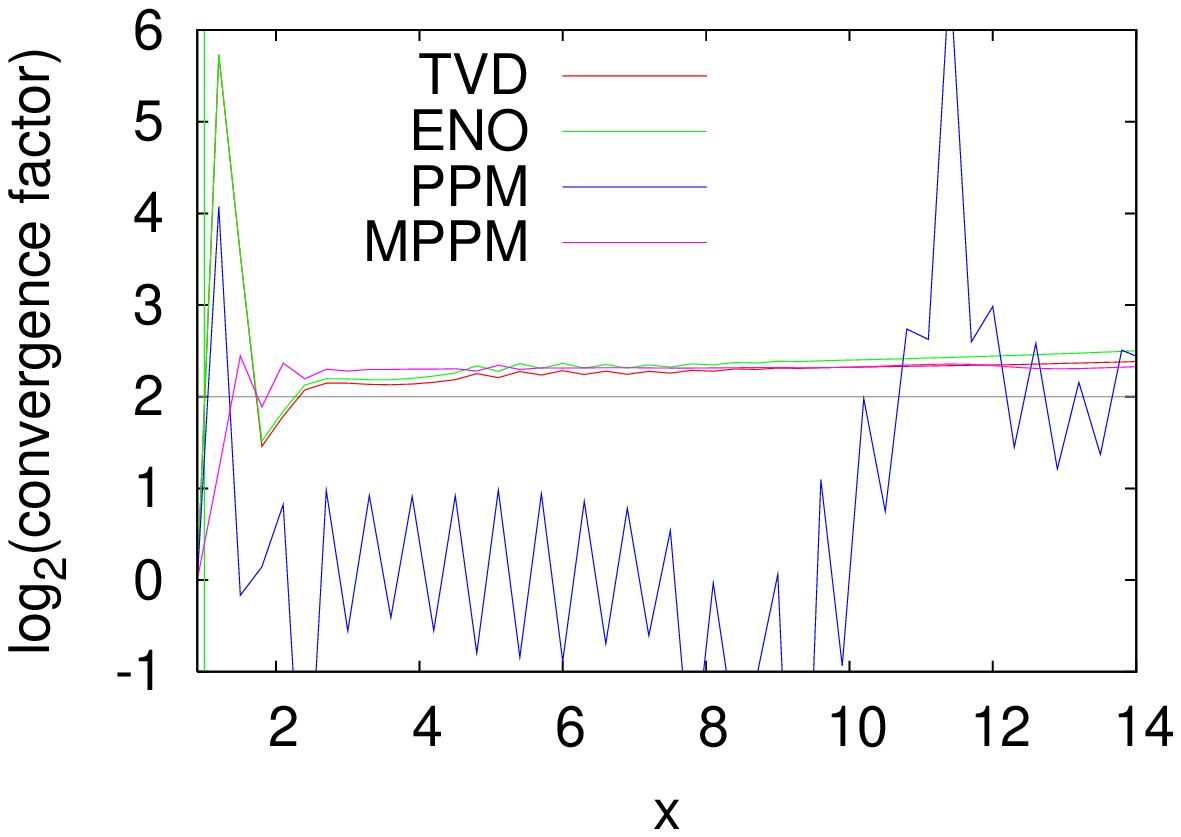}
    \caption{Michel solution compared with the numerical evolution for
      the ENO method. All other methods look very similar in the
      physical quantities. The graphs show the values for the $x$
      component of the velocity $v_x$, the density $\rho$, and the
      pressure $p$ of the fluid.  Each quantity is evolved until time
      $t=40$ (in units of the background black hole mass), and the
      numerical values are compared with the exact solution (the solid
      line in the graphs).  The $x$ axis is the radial distance
      expressed in black hole units.  The excision region is delimited
      by the vertical, dotted line.  The last graph shows the
      convergence factor of all reconstruction methods, where
      the expected second--order
      convergence is seen for all methods except for PPM.}
    \label{fig:michell}
  \end{center}
\end{figure*}

The first three plots in Fig.~\ref{fig:michell} show a part of the
results of a series of runs made to test the Michel solution. In these
plots the ENO reconstruction method is used, but the graphs look very
similar for TVD, PPM and MPPM. We set the Michel solution to be our
initial data, then we evolve it and verify its accuracy.  In the
plot, the exact solution is compared with the computed solution at
time $t=40$ (in units of the background black hole mass). Using a logarithmic
plot we show the results for the density, $x$-velocity and pressure.
The two sets of data are in very good
agreement, as expected from the staticity of the solution.  It is
possible to notice a slight discrepancy in the first grid point next
to the excision region (dotted line).  This discrepancy is related to
the lower accuracy which we have there, visually amplified on the plot
by the choice of logarithmic scale. This is the case for all
reconstruction methods.

The fourth plot in Fig.~\ref{fig:michell} shows instead the
convergence factor obtained for all the reconstruction methods on the
same set of simulations. As evident in the figure, PPM does not show
convergence, while the other reconstruction methods show the expected
convergence factor of approximately two. We will discuss the possible reason
for that later.

\subsection{Neutron Star collapse}
\label{sec:NScollapse}

As the next test we consider the collapse of a neutron star, in a
simulation in which both the spacetime and the matter are evolved.
This test is considerably more difficult than previous tests,
requiring both hydrodynamic fields and spacetime to be consistently
excised, and requiring a dynamic excision region.

As in~\cite{Font02c} a spherically symmetric polytrope with $K=100$,
$\Gamma=2$ and a central density of $\rho_c=8\times 10^{-3} \approx
4.95\times10^{15} \, \textrm{g} \, \textrm{cm}^{-3}$ is used.  The
collapse was induced by lowering $K$ by $2\%$ and rescaling the matter
variables to enforce the solution of the constraints.  Note that
because this satisfies the constraints and the spacetime is not
changed, the ADM mass of the system is also not changed by this.

Due to the symmetry of the problem we only evolve one octant of the
grid. The spacetime is evolved using the NOK formulation and the
excision methods described in~\cite{Alcubierre00a,Alcubierre01a}. The
apparent horizon, if it can be found, is located using the horizon
finder described in~\cite{Thornburg2003:AH-finding}. The gauge
conditions used are ``1+log'' slicing for the lapse and the
gamma-driver condition given in equation (45) of~\cite{Alcubierre02a}.
The event horizon is located using the code described
in~\cite{Diener03a}.

\begin{figure*}[htbp]
  \begin{center}
    \includegraphics*[width=12cm]{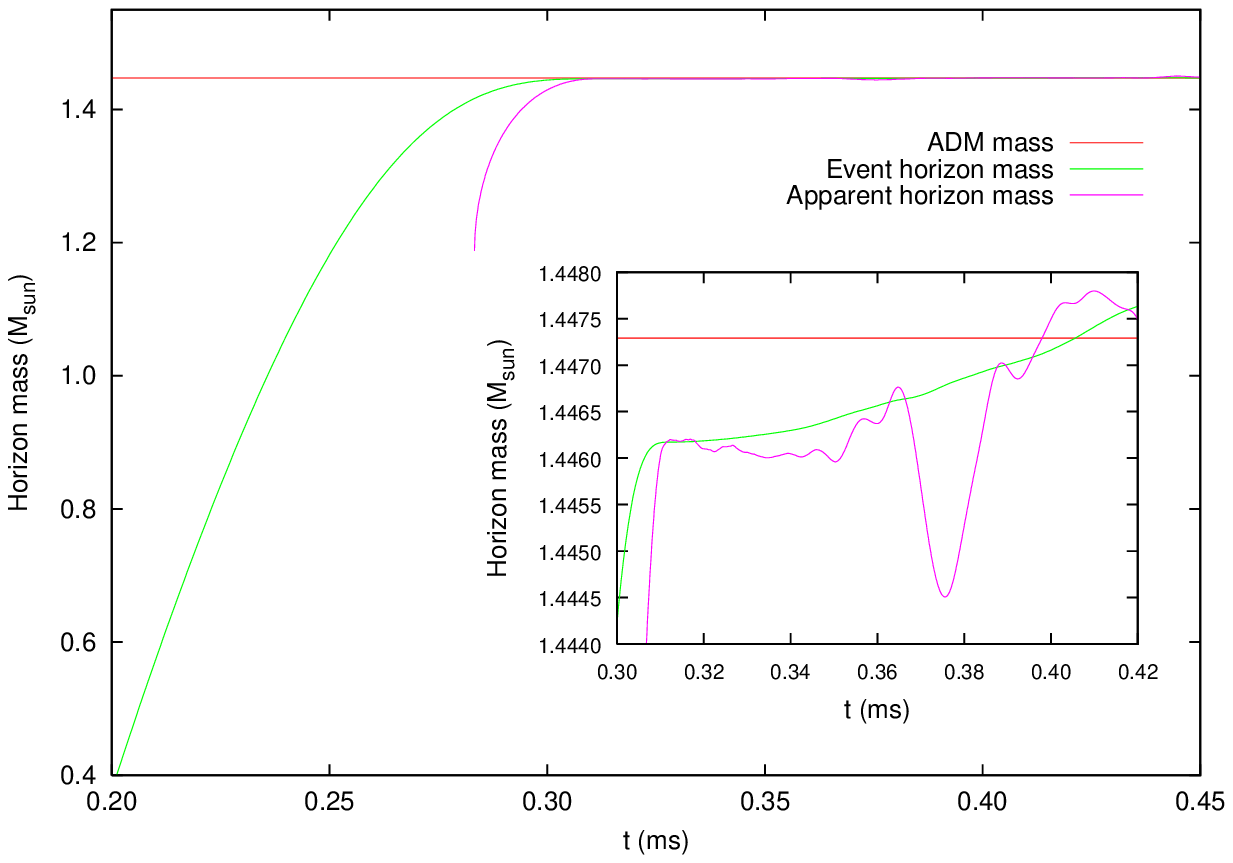}
    \caption{Horizon masses for the collapse of a spherically
      symmetric neutron star. At early times the evolution is stable
      and accurate. The neutron star collapses smoothly into the
      horizon and the matter is excised. The horizon masses as
      measured by apparent and event horizon finders are very close to
      the total ADM mass of the initial spacetime, as expected. At
      late times instabilities lead to inaccuracies, as seen in the
      inset, before the simulation fails.}
    \label{fig:ahmass}
  \end{center}
\end{figure*}
In Fig.~\ref{fig:ahmass} we show the mass of the event and apparent
horizons of the black hole found by measuring their surface areas and using
$M=\left( A / 16 \pi \right)^{1/2}$ and the ADM mass computed by the
inital data solver for comparison.  At early times the expected
results are seen: the mass of the apparent horizon is less than the
mass of the event horizon which is less than the total ADM mass of the
spacetime. At late times large violations of the Hamiltonian
constraint develop, errors in the determination of the
horizon become large and lead eventually to the crash of the code at
around $t=0.455$ms.
Without excision the code crashes shortly after formation of the
apparent horizon.

We believe that the origin of these instabilities lies in the
spacetime excision technique, and not the hydrodynamic excision
described here. This can be seen much more clearly in another test,
where we used a slowly rotating neutron star
(model D1 of~\cite{Baiotti04})
and excise the spacetime variables
and the matter variables at different times and compare the evolution
of the 2-norm of the Hamiltonian constraint in the two cases.  Note
that it is not possible to first excise the spacetime variables and
later the hydrodynamics because the hydrodynamics code is very
sensitive to errors in the spacetime. Therefore we always first excise the
hydrodynamical variables and then switch on the excision of the
spacetime at a later time. We excise the hydrodynamic variables
shortly after an apparent horizon is found (which in this case is at a
{\it coordinate} time of $0.546$ms) and vary the coordinate time,
$t_{\text{ex}}$, when the spacetime is first excised.

\begin{figure*}[htbp]
  \begin{center}
    \includegraphics*[width=12cm]{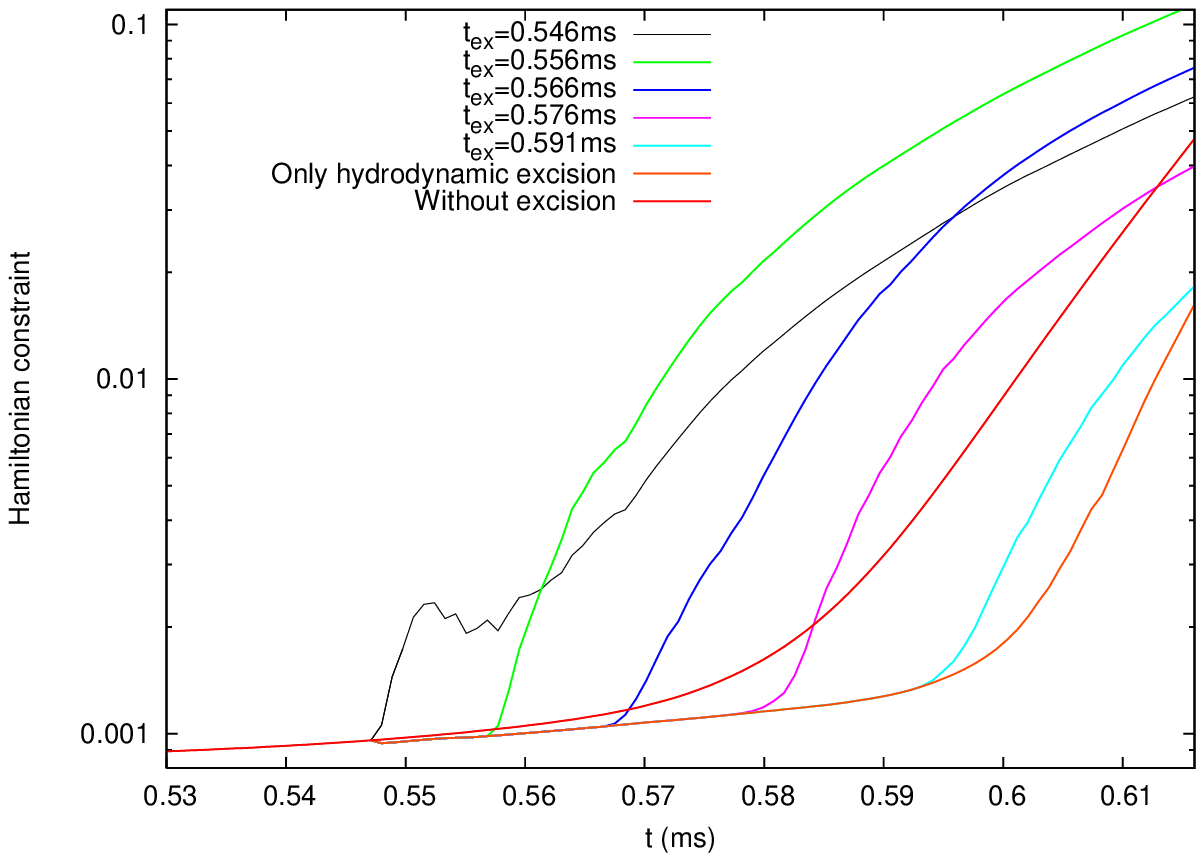}
    \caption{The Hamiltonian constraint violation over time, plotted
      for different starting times $t_{\text{ex}}$ of the spacetime
      excision. In every case (except the reference case where no
      excision is used) the hydrodynamical variables are excised as
      soon as the apparent horizon is found at $t=0.547$ms.
      Instabilities at the excision boundary are believed to be the
      cause of the eventual failure of the simulation. Here it is
      clear that when the excision boundary condition is applied to
      the spacetime variables there is an immediate exponential growth
      in the constraint violation. In contrast, no growth is seen at
      the time when the hydrodynamical variables are excised. The
      instabilities which are indicated by the exponential growth of
      the constraint violations are clearly triggered by the spacetime
      excision.}
    \label{fig:diff_ex_times}
  \end{center}
\end{figure*}

Figure \ref{fig:diff_ex_times} shows the Hamiltonian constraint
violation for simulations where the excision time $t_{\text{ex}}$ is
varied. For comparison, a simulation where neither the hydrodynamical
variables nor the spacetime variables were excised, and a simulation
where only the hydrodynamical variables are excised, are also shown.

Where the hydrodynamical or spacetime variables are excised the
Hamiltonian constraint violation inside the excised region is not
taken into account. At the point where either set of variables is
excised the constraints are clearly meaningless within the excision
region. We expect this to have no effect on the exterior spacetime as
demonstrated numerically in,
e.g.,~\cite{Alcubierre2003:BBH0-excision}. The constraint violations in
the exterior spacetime clearly are meaningful, and we will be most
interested in the behaviour shortly after either set of
variables is excised

It is clear that shortly after the time when the spacetime variables
are excised there is a sudden exponential increase in the violation of
the constraints. This does not happen at the time when the excision of
the hydrodynamical variables first takes place. We believe this
indicates that the cause of the instabilities encountered is the
method used to excise the spacetime variables.  Whilst we cannot rule
out instabilities from the hydrodynamical excision, the simulation
where only the hydrodynamical variables are excised seems to lose
accuracy due to classical slice stretching effects and not due to the
high frequency oscillations which are instead observed when excising
the spacetime variables.


\section{Conclusion}
\label{sec:conclusion}

Long term evolutions involving black holes and matter will require
excision boundary conditions for the matter fields. For hydrodynamics,
where the fluid will generically form shock waves, the results of Lax
and Wendroff~\cite{Lax60} and Hou and Lefloch~\cite{Hou94} imply that
a conservative total-variation stable scheme must be used, which
presently means a HRSC scheme. Although errors in the hydrodynamics at
the excision boundary should not propagate outside of the horizon and
so should have no effect on the physics, such inconsistencies may lead
to numerical instabilities.

In this paper we have shown a simple method of providing a consistent
excision boundary condition for systems of equations in conservation
law form evolved with HRSC schemes. Although we have only applied this
technique to the Valencia formulation of relativistic
hydrodynamics~\cite{Marti91,Banyuls97,Ibanez01} we expect the results
to apply to any system where the conservation law form can be applied.
The method is simple and works on a broad range of HRSC schemes.

The tests of section~\ref{sec:tests} show that the method works for
shocks, steady state cases, and dynamical evolutions. The shock tests
are particularly simple and enlightening, showing that an incorrect
boundary condition may lead to inconsistency or numerical
instabilities. The Michel solution test shows that our boundary
conditions are also suited to long term steady state evolutions with
strong gradients.

The dynamical collapse of a spherically symmetric TOV star shows the
advantages and current limitations of our methods. The collapse to a
black hole is accurately followed long past the formation of the
horizon, the mass of which is computed to high accuracy. However, the
simulation does not continue indefinitely, eventually failing due to a
numerical instability. Simulations including rotation show the same
features and, by varying the initial time at which excision is
applied, there is evidence that the cause of the instabilities is the
excision method applied to the spacetime variables, not the
hydrodynamical method that is the focus of this paper. However we want
to emphasise that although we believe that the simple lego excision
boundary condition is the cause of the instabilities, the excision of
the spacetime variables is nevertheless improving the length of the
simlulations in the cases described here.

It has been suggested that HRSC methods may improve spacetime
evolutions for certain hyperbolic formulations of Einsteins
Equations~(e.g., \cite{Bona94b,Bardeen02,Bona04b}).  If such a method
were used, excision boundary conditions such as those presented in
this paper should be easy to adapt. There are also other approaches to
applying excision boundary conditions that show great promise (e.g.,
\cite{Calabrese2003:excision-and-summation-by-parts,Calabrese:2003vx,Thornburg2004:multipatch-BH-excision}).
It is, however, not clear if these methods will work in situations
where fundamental variables may become discontinuous. In such cases
the methods presented here could then be used for a matter field where
the evolution equations could be written in conservation law form, and
any alternative method could be used for the spacetime.

At no point here we have addressed the question of the well-posedness
of the system used, or any analytical proof of the stability of
our method. The well-posedness of the simple Euler equations with
arbitrary initial data in more than one dimension in the absence of
gravity is not yet established; the most relevant current result
(\cite{Barnes04}) considers the one dimensional Euler equations on a
plane-symmetric Gowdy spacetime. Given the current state of knowledge,
and the successes in extending one-dimensional schemes that are known
to be stable to more than one dimension, we expect that methods such
as the one given here will be stable but are unable to give a proof at
any level.


\acknowledgments

We want to thank all people involved in the effort of building the
Whisky code, Jonathan Thornburg for his apparent horizon finder, Peter
Diener for his event horizon finder and all people in the Cactus-code
team for producing an efficient infrastructure for our work.  We are
also very grateful to Luciano Rezzolla and the people at the AEI for
discussions and want to thank Luca Baiotti, Denis Pollney, Luciano Rezzolla,
Ed Seidel and Jonathan Thornburg for proof-reading the manuscript.
The simulations were performed on the peyote cluster at
the AEI and an Origin 300 at Portsmouth, funded by a UK SRIF grant. IH
was partially supported by PPARC grant PPA/G/S/2002/00531. AN was partly
supported by NASA grant NNG04GL37G to the University of Texas at Austin.


\bibliographystyle{apsrev}

\bibliography{bibtex/references}


\end{document}